\documentclass[aps,prl,twocolumn,showpacs,preprintnumbers,amsmath,amssymb,superscriptaddress]{revtex4-1}

\usepackage{graphicx}
\usepackage{xspace}
\usepackage{siunitx}
\usepackage[usenames]{color}
\usepackage{comment}
\usepackage{bigdelim}
\usepackage{amssymb}
\usepackage{amsmath}
\usepackage{array}
\usepackage{multirow}
\usepackage{tabularx}
\usepackage{dcolumn}
\usepackage{booktabs,dcolumn}
\usepackage{bm}
\usepackage{float}
\usepackage{fancyhdr}
\usepackage{afterpage}
\usepackage{url}
\usepackage{bm}
\usepackage{epstopdf}

\usepackage{enumitem}
\usepackage{setspace}
\usepackage{color}
\usepackage{hyperref}
\hypersetup{colorlinks,urlcolor=blue,citecolor=black}
\usepackage{listings}
\usepackage{geometry}
\usepackage[dvipsnames]{xcolor}

\graphicspath{{./figs/}}

\newcommand{\dd}{\mathrm{d}}

\newcommand{\eqs}[1]{Eqs.~\ref{eq:#1}}

\newcommand{\abs}[1]{\left|#1\right|}
\newcommand{\etal}{{\it et al.~}}

\newcommand{\nba}[1]{} 

\newcommand{\qmax}{\ensuremath{Q_{\mathrm{max}}}\xspace}
\newcommand{\qmin}{\ensuremath{Q_{\mathrm{min}}}\xspace}

\newcommand{\pdfgui}{\textsc{PDFgui}\xspace}

\begin{document}

\makeatletter
\@ifundefined{textcolor}{}
{%
 \definecolor{BLACK}{gray}{0}
 \definecolor{WHITE}{gray}{1}
 \definecolor{RED}{rgb}{1,0,0}
 \definecolor{GREEN}{rgb}{0,1,0}
 \definecolor{BLUE}{rgb}{0,0,1}
 \definecolor{CYAN}{cmyk}{1,0,0,0}
 \definecolor{MAGENTA}{cmyk}{0,1,0,0}
 \definecolor{YELLOW}{cmyk}{0,0,1,0}
}


\makeatother



\preprint{preprint(\today)}

\title{Possible topologically non-trivial superconducting order parameter in type-II Weyl semimetal $T_{d}$-MoTe$_{2}$}

\author{Z.~Guguchia}
\email{zg2268@columbia.edu} \affiliation{Department of Physics, Columbia University, New York, NY 10027, USA}
\author{F.v.~Rohr}
\affiliation{Department of Chemistry, Princeton University, Princeton, New Jersey 08544, USA}

\author{Z.~Shermadini}
\affiliation{Laboratory for Muon Spin Spectroscopy, Paul Scherrer Institute, CH-5232
Villigen PSI, Switzerland}

\author{A.T.~Lee}
\affiliation{Department of Applied Physics and Applied Mathematics, Columbia University, New York, NY 10027, USA}

\author{S. Banerjee}
\affiliation{Department of Applied Physics and Applied Mathematics, Columbia University, New York, NY 10027, USA}

\author{A.R.~Wieteska}
\affiliation{Department of Physics, Columbia University, New York, NY 10027, USA}

\author{C.A.~Marianetti}
\affiliation{Department of Applied Physics and Applied Mathematics, Columbia University, New York, NY 10027, USA}

\author{H.~Luetkens}
\affiliation{Laboratory for Muon Spin Spectroscopy, Paul Scherrer Institute, CH-5232
Villigen PSI, Switzerland}

\author{Z. Gong}
\affiliation{Department of Physics, Columbia University, New York, NY 10027, USA}

\author{B.A. Frandsen}
\affiliation{Department of Physics, University of California, Berkeley, California 94720, USA}

\author{S.C. Cheung}
\affiliation{Department of Physics, Columbia University, New York, NY 10027, USA}

\author{C.~Baines}
\affiliation{Laboratory for Muon Spin Spectroscopy, Paul Scherrer Institute, CH-5232
Villigen PSI, Switzerland}

\author{A.~Shengelaya}
\affiliation{Department of Physics, Tbilisi State University, Chavchavadze 3, GE-0128 Tbilisi, Georgia}
\affiliation{Andronikashvili Institute of Physics of I.Javakhishvili Tbilisi State University,
Tamarashvili str. 6, 0177 Tbilisi, Georgia}

\author{A.N.~Pasupathy}
\affiliation{Department of Physics, Columbia University, New York, NY 10027, USA}

\author{E.~Morenzoni}
\affiliation{Laboratory for Muon Spin Spectroscopy, Paul Scherrer Institute, CH-5232
Villigen PSI, Switzerland}

\author{S.J.L.~Billinge}
\affiliation{Department of Applied Physics and Applied Mathematics, Columbia University, New York, NY 10027, USA}
\affiliation{Condensed Matter Physics and Materials Science Department,
Brookhaven National Laboratory, Upton, NY 11973, USA}

\author{A.~Amato}
\affiliation{Laboratory for Muon Spin Spectroscopy, Paul Scherrer Institute, CH-5232
Villigen PSI, Switzerland}

\author{R.J. Cava}
\affiliation{Department of Chemistry, Princeton University, Princeton, New Jersey 08544, USA}

\author{R.~Khasanov}
\affiliation{Laboratory for Muon Spin Spectroscopy, Paul Scherrer Institute, CH-5232
Villigen PSI, Switzerland}

\author{Y.J.~Uemura}
\affiliation{Department of Physics, Columbia University, New York, NY 10027, USA}


\maketitle

\textbf{MoTe$_2$, with the orthorhombic $T_{d}$ phase, is a new type (type-II) of Weyl semimetal \cite{Soluyanov,SunY,WangZ,KourtisS,KDeng,NXu,Kaminski,Tamai1}, where the Weyl Fermions emerge at the boundary between electron and hole pockets.  
Non-saturating magnetoresistance (MR) \cite{Xu,Ali1,Balicas1,ZhuZ}, and superconductivity \cite{PanXC,KangD,QiCava}  were also observed in $T_{d}$-MoTe$_{2}$. Understanding the superconductivity in $T_{d}$-MoTe$_{2}$, which was proposed to be topologically non-trivial, is of eminent interest. Here, we report high-pressure ($p_{\rm max}$ = 1.3 GPa) muon spin rotation experiments on the temperature-dependent magnetic penetration depth $\lambda\left(T\right)$ in $T_{d}$-MoTe$_{2}$. 
A substantial increase of the superfluid density $n_{s}/m^{*}$ and a linear scaling with $T_{c}$ is observed under pressure. Moreover, the superconducting order parameter in $T_{d}$-MoTe$_{2}$ is determined to be two gap ($s+s$)-wave symmetric. We also excluded time reversal symmetry breaking in the SC state with sensitive zero-field ${\mu}$SR experiments. Considering the previous report \cite{Balicas1} on the strong  suppression of $T_{\rm c}$  in MoTe$_{2}$ by disorder, we suggest that $s^{+-}$ (topological order parameter) state is more likely to be realized in MoTe$_{2}$ than the $s^{++}$ (trivial) state. Should $s^{+-}$ be the SC gap symmetry, the $T_{d}$-MoTe$_{2}$ is, to our knowledge, the first known example of a time reversal invariant topological (Weyl) superconductor.}

 Transition metal dichalcogenides (TMDs) have attracted a lot of attention due to their fascinating physics and
promising potential applications \cite{Klemm, Qian, Xu, Morosan, LiY, Zhang}. TMDs share the
same formula, MX$_{2}$, where M is a transition metal (for
example, Mo or W) and X is a chalcogenide atom (S, Se and Te). These compounds typically crystallize in a group of related structure types, including 2H-, 1T-, 1T$^{'}$- and $T_{d}$-type lattices \cite{Clarke,Puotinen,Zandt,Brown}. 
The 2H-MoTe$_{2}$ is semiconducting, while the 1T$^{'}$- and $T_{d}$-MoTe$_{2}$ are
semimetallic and exhibit pseudo-hexagonal layers with zig-zag metal chains. 
$T_{d}$-MoTe$_2$ and WTe$_2$ materials are considered as a type-II Weyl semimetal \cite{Soluyanov,SunY}.
The Fermi surfaces in a type-II Weyl semimetal consist of a pair of electron- and hole- pockets touching at the Weyl node, rather than at the point-like Fermi surface in traditional WSM (Type I, e.g. TaAs) systems. 
The Weyl fermions can also be different from each other in terms of their symmetry-breaking origin. In general, they can arise by breaking either the space-inversion (SIS) or time-reversal symmetry (TRS). 
At the same time, however, at least one of them must be preserved to realize a bulk superconducting state \cite{Hosur,Ando,Grushin}. These different categories of Weyl semimetals are expected to exhibit distinct topological properties. 

 MoTe$_{2}$ exhibits a monoclinic 1T$^{'}$-orthorhombic $T_{d}$ structural phase transition at $T_{\rm S}$ ${\sim}$ 250 K \cite{QiCava}. We note that the 1T$^{'}$ structure exhibits the inversion symmetric space group $P$2$_{1}$/$m$, while the $T_{d}$ phase belongs to the non-centrosymmetric space group ${Pmn}$2$_{1}$. 
Weyl fermions occur in the $T_{d}$ phase where the inversion symmetry is broken \cite{Soluyanov}.
The  evidence for the low temperature $T_{d}$ structure in our MoTe$_{2}$ sample is provided by X-ray
pair distribution function (PDF) measurements (see Supplementary section II and Supplementary Figures 7-9). Recently, a surface state connecting bulk hole pockets and bulk electron ones and which also dominates the signal at the Fermi level, was observed by angle-resolved photoemission (ARPES) measurements\cite{Liang}. In addition, high field quantum oscillation study of the MR for $T_{d}$-MoTe$_{2}$, revealed a nontrivial ${\pi}$ Berry's phase in $T_{d}$-MoTe$_{2}$, which is a distinguished feature of surface states \cite{Luo}. Additionally, in the Mo$_{x}$W$_{1-x}$Te$_{2}$, experimental signatures of the predicted topological connection between the Weyl bulk states and Fermi arc surface states, which are a unique property of a Weyl semimetal were also reported \cite{Zheng}.

\begin{figure}[t!]
\includegraphics[width=1.0\linewidth]{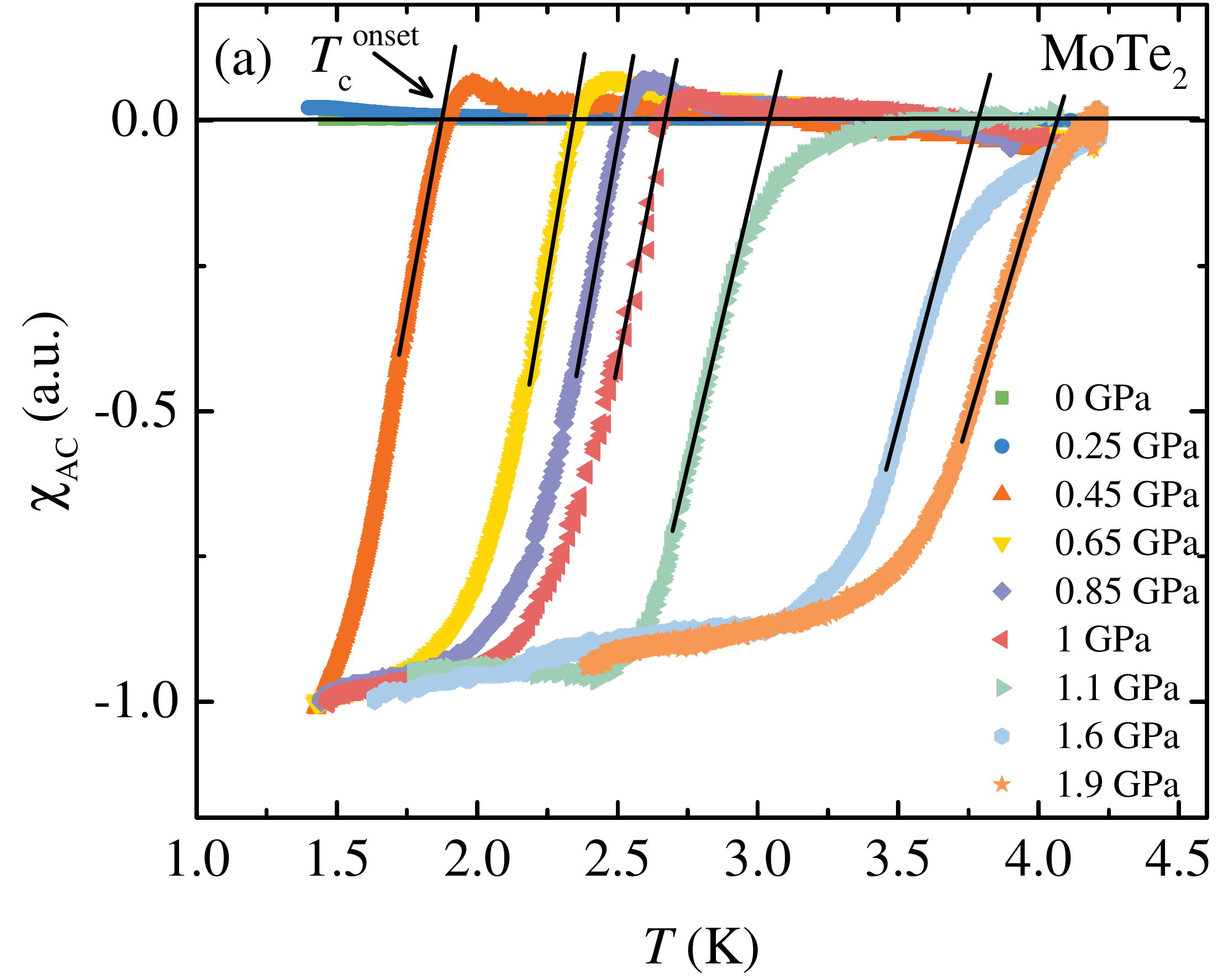}
\includegraphics[width=1.0\linewidth]{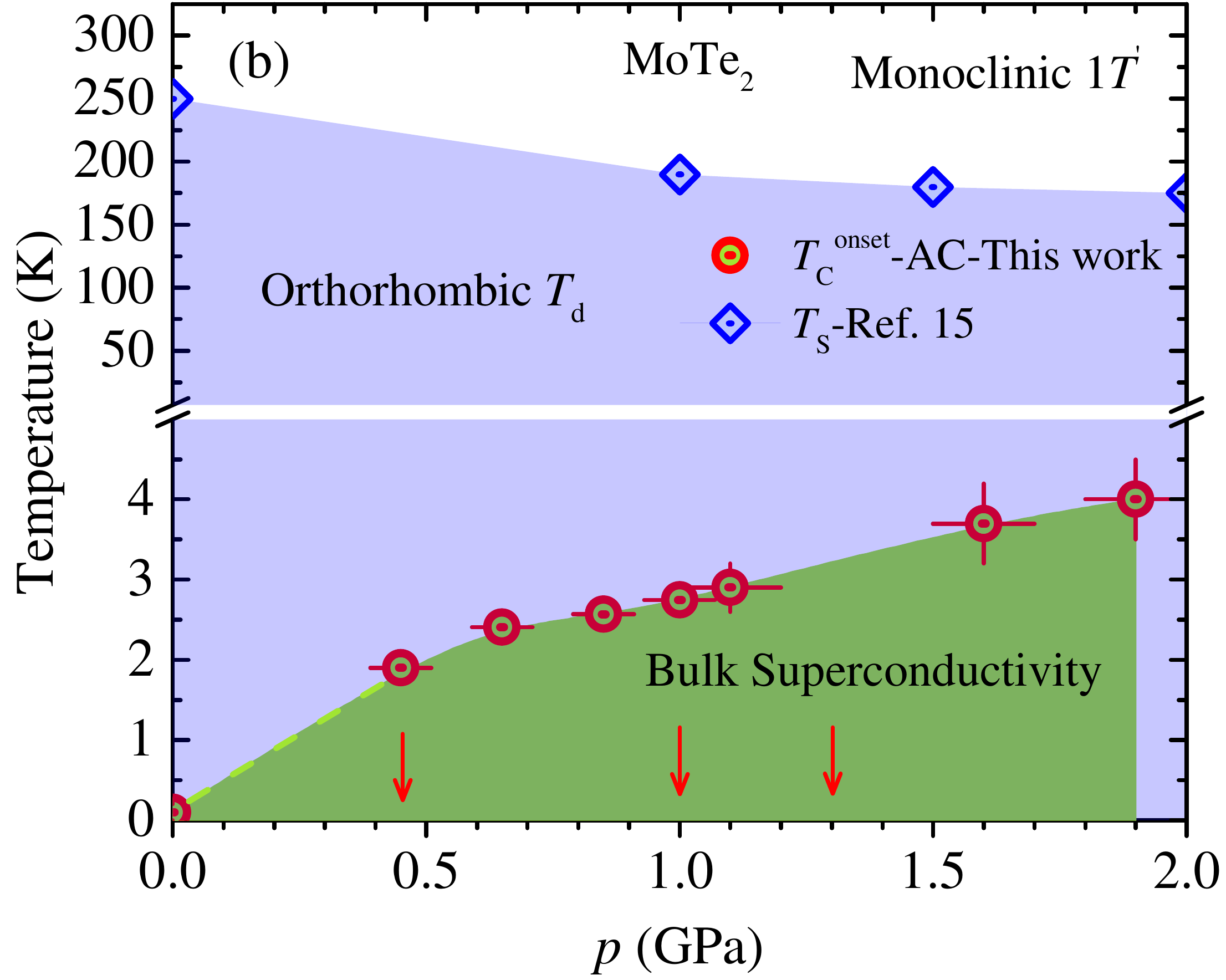}
\vspace{-0.8cm}
\caption{ (Color online) \textbf{AC susceptibility as a function of temperature and pressure MoTe$_{2}$.} 
(a) Temperature dependence of the AC susceptibility $\chi_{\rm AC}$ for the polycrystalline sample of MoTe$_{2}$, measured at ambient and at various applied hydrostatic pressures up to $p$ ${\simeq}$ 1 GPa. The arrows denote the superconducting transition temperature $T_{\rm c}$. (b) Pressure dependence of $T_{\rm c}$ (this work) and the structural phase transition temperature $T_{\rm S}$ (ref.  \cite{QiCava}). Arrows mark the pressures, at which the $T$-dependence of the penetration depth was measured.}
\label{fig1}
\end{figure}
\begin{figure*}[t!]
\centering
\includegraphics[width=1.0\linewidth]{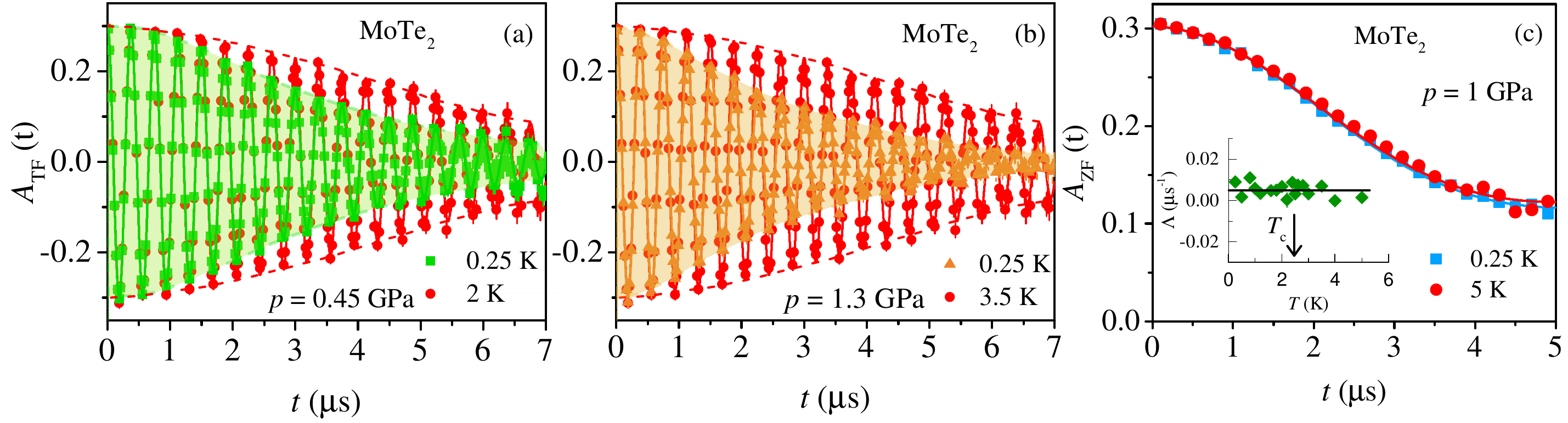}
\vspace{-0.8cm}
\caption{ (Color online) \textbf{Transverse-field (TF) and zero-field (ZF) ${\mu}$SR time spectra for MoTe$_{2}$.} 
The TF spectra are obtained above and below $T_{\rm c}$ (after field cooling the sample from above $T_{\rm c}$): (a)  $p$ = 0.45 GPa and (b) $p$ = 1.3 GPa. The solid lines in panel a and b represent fits to the data by means of Eq.~1. The dashed lines are the guides to the eyes. (c) ZF ${\mu}$SR time spectra for MoTe$_{2}$ recorded above and below $T_{\rm c}$. The line represents
the fit to the data of a standard Kubo-Toyabe depolarization function \cite{Toyabe}, reflecting
the field distribution at the muon site created by the nuclear moments.}
\label{fig2}
\end{figure*}

 $T_{d}$-MoTe$_{2}$ represents a rare example of a material with both a topologically non-trivial band structure and superconductivity. 
$T_{d}$-MoTe$_{2}$ is superconducting with $T_{c}$ ${\simeq}$ 0.1 K \cite{QiCava}. The application of small pressure \cite{QiCava} or the substitution of Te ion by S \cite{Chen} can dramatically enhance $T_{c}$. $T_{d}$-MoTe$_{2}$ is believed to be a promising candidate of topological superconductor (TSC) in the bulk materials. TSCs are special families of those materials with unique electronic states, a full pairing gap in the bulk and gapless surface states consisting of Majorana fermions (MFs) \cite{Hosur,Ando,Grushin}.
Due to their scientific importance and potential applications in quantum computing, MFs have attracted lots of attention recently. 
In general, topological superfluidity, and superconductivity, are well-established phenomena in condensed matter systems. The A-phase of superfluid helium-3 constitutes an example of a charge neutral topological superfluid, whereas Sr$_{2}$RuO$_{4}$ \cite{LukeTRS}, is generally believed to be topological TRS breaking superconductor. However, an example of a TRS invariant topological superconductor \cite{Hosur,Ando} is outstanding and $T_{d}$-MoTe$_{2}$ may be a candidate material for this category. So far, pressure-dependent critical temperatures and fields are the only known properties of the superconducting state of $T_{d}$-MoTe$_{2}$. Thus, thorough exploration of superconductivity in $T_{d}$-MoTe$_{2}$ from both experimental and theoretical perspectives are required.

To further explore superconductivity in $T_{d}$-MoTe$_{2}$ and its topological nature, it is extremely important to measure 
the superconducting order parameter of $T_{d}$-MoTe$_{2}$ on the microscopic level
through measurements of the bulk properties. Thus, we concentrate on the high pressure \cite{GuguchiaPressure, Andreica, MaisuradzePC, GuguchiaNature} muon spin relaxation/rotation (${\mu}$SR)  measurements of the magnetic penetration depth $\lambda$ in $T_{d}$-MoTe$_{2}$, which is one of the fundamental parameters of a superconductor, since it is related to the superfluid density $n_{s}$ via 1/${\lambda}^{2}$ = $\mu_{0}$$e^{2}$$n_{s}/m^{*}$ (where $m^{*}$ is the effective
mass). Most importantly, the temperature dependence of ${\lambda}$
is particularly sensitive to the topology of the SC gap: while in a
fully gapped superconductor, $\Delta\lambda^{-2}\left(T\right)\equiv\lambda^{-2}\left(0\right)-\lambda^{-2}\left(T\right)$
vanishes exponentially at low $T$, in a nodal SC it vanishes as a
power of $T$. The muon-spin rotation (${\mu}$SR) technique provides
a powerful tool to measure ${\lambda}$ in the vortex state of type II superconductors in the bulk of the sample, in contrast to many techniques that probe  ${\lambda}$ only near the surface \cite{Sonier}. 
Details are provided in the Methods section. Zero-field ${\mu}$SR is very powerful for detecting 
a spontaneous magnetic field due to TRS breaking in exotic superconductors, because internal magnetic fields as small as 0.1 G are detected in measurements without applying external magnetic fields. 
High-pressure AC susceptibility experiments were also carried out on the same polycrystalline sample of $T_{d}$-MoTe$_{2}$ using exactly the same cell as the one used for ${\mu}$SR experiments.

 Figure 1a shows the temperature dependence of the AC-susceptibility $\chi_{\rm AC}$ of $T_{d}$-MoTe$_{2}$ in the temperature range between 1.4 K and 4.2 K for selected hydrostatic pressures up to $p$ = 1.9 GPa. 
Strong diamagnetic response and sharp SC transitions are observed under pressure (Fig. 1), pointing to the high quality of the sample and providing evidence for bulk superconductivity in MoTe$_{2}$ \cite{QiCava}. The pressure dependence of $T_{{\rm c}}$ is shown in Fig. 1b. $T_{{\rm c}}$ increases with increasing pressure and reaches a critical temperature $T_{\rm c}$ ${\simeq}$ 4 K at the maximum applied pressure of $p$ = 1.9 GPa in susceptibility experiments. The substantial increase of $T_{\rm c}$ from  $T_{\rm c}$ ${\simeq}$ 0.1 K at ambient pressure to $T_{\rm c}$ ${\simeq}$ 4 K at moderate pressures in MoTe$_{2}$ was considered as a manifestation of its topologically nontrivial electronic structure. Note that a strong pressure-induced enhancement of $T_{\rm c}$ has also been observed in topological superconductors such as Bi$_{2}$Te$_{3}$ \cite{JLZhang} and Bi$_{2}$Se$_{3}$ \cite{Kirshenbaum}. In Fig. 1b, the monoclinic 1T$^{'}$-orthorhombic $T_{d}$ structural phase transition  temperature $T_{\rm S}$ \cite{QiCava} is also shown as a function of pressure demonstrating that in the investigated pressure range $p$ = 0 - 1.9 GPa, the system MoTe$_{2}$ exhibits the orthorhombic $T_{d}$ structure. Moreover, in the investigated pressure region, the system MoTe$_{2}$ was also confirmed by DFT calculations \cite{QiCava} to be a Weyl semimetal with a band structure around the Fermi level, which is extremely sensitive to changes in the lattice constants.

  Figures \ref{fig2}a and \ref{fig2}b exhibit the transverse-field (TF) ${\mu}$SR-time spectra for MoTe$_{2}$, measured at $p$ = 0.45 GPa and maximum applied pressure $p$ = 1.3 GPa, respectively. The spectra above (2 K, 3.5 K) and below (0.25 K) the SC transition temperature $T_{{\rm c}}$ are shown. Above $T_{{\rm c}}$ the oscillations show a small relaxation due to the random local fields from the nuclear magnetic moments. Below $T_{{\rm c}}$ the relaxation rate strongly increases with decreasing temperature due to the presence of a nonuniform local magnetic field distribution as a result of the
formation of a flux-line lattice (FLL) in the SC state. Magnetism, if present in the samples, may enhance the muon depolarization rate and falsify the interpretation of the TF-${\mu}$SR results. Therefore, we have carried out ZF-${\mu}$SR experiments above and below $T_{{\rm c}}$ to search for magnetism (static or fluctuating) in MoTe$_{2}$. As shown in \ref{fig2}c no sign of either static or fluctuating magnetism could be detected in ZF time spectra down to 0.25 K. The spectra are well described by a standard Kubo-Toyabe depolarization function \cite{Toyabe}, reflecting the field distribution at the muon site created by the nuclear moments. Moreover, no change in ZF-${\mu}$SR relaxation rate (see the inset of Fig. \ref{fig2}c) across $T_{c}$ was observed, pointing to the absence of any spontaneous magnetic fields associated with a TRS \cite{LukeTRS,HillierTRS,BiswasTRS} breaking pairing state in MoTe$_{2}$.

\begin{figure}[b!]
\includegraphics[width=0.98\linewidth]{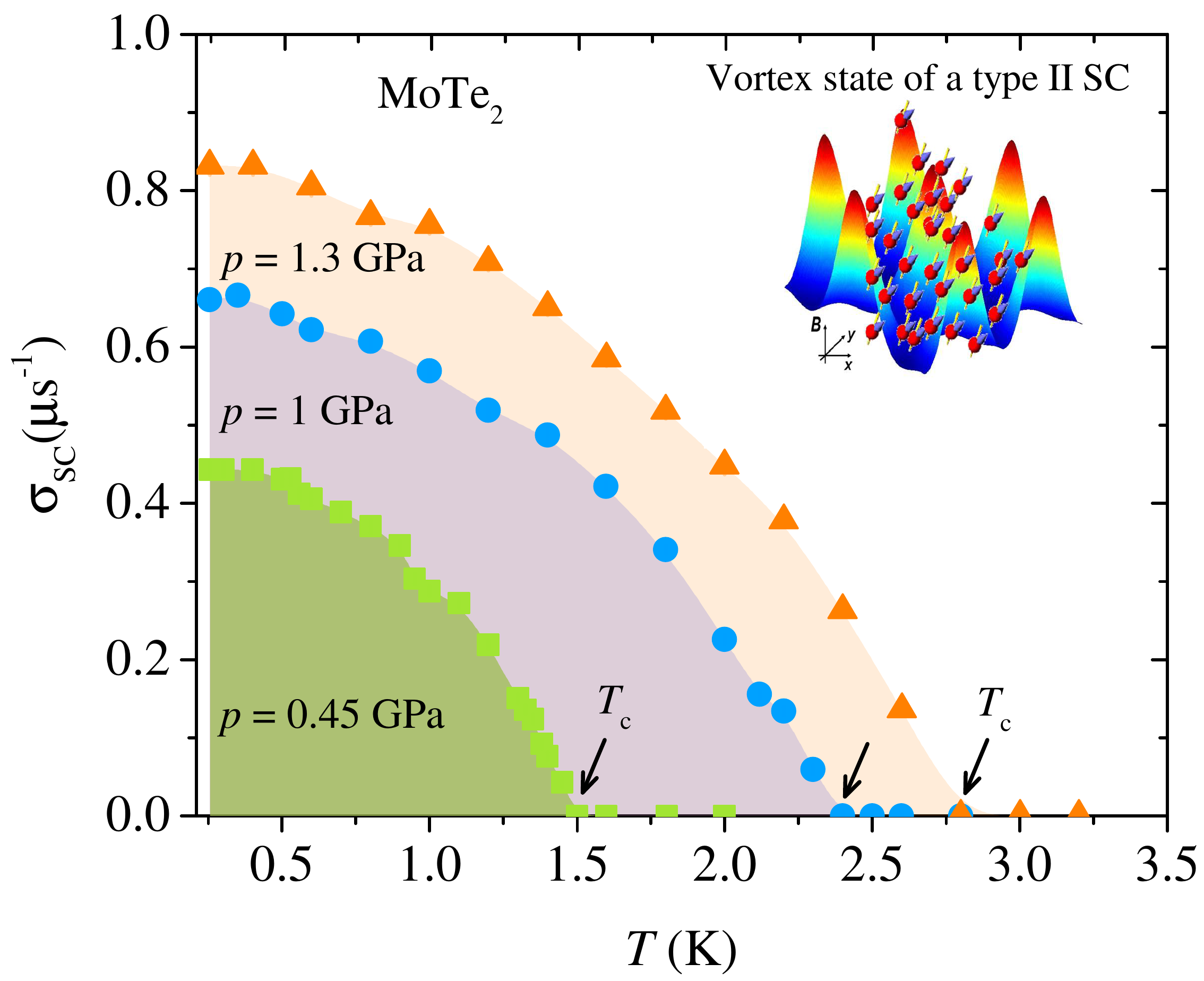}
\vspace{-0.4cm}
\caption{(Color online)  \textbf{Superconducting muon spin depolarization rate ${\sigma}_{\rm sc}$ for MoTe$_{2}$.} Temperature dependence of ${\sigma}_{\rm sc}$($T$), measured at various hydrostatic pressures in an applied magnetic field of ${\mu}_{\rm 0}H = 20$~mT. The arrows mark the $T_{c}$ values. Inset illustrates how muons, as local probes, sense the inhomogeneous field distribution in the vortex state of Type II superconductor.}
\label{fig3}
\end{figure}
\begin{figure*}[t!]
\centering
\includegraphics[width=1.0\linewidth]{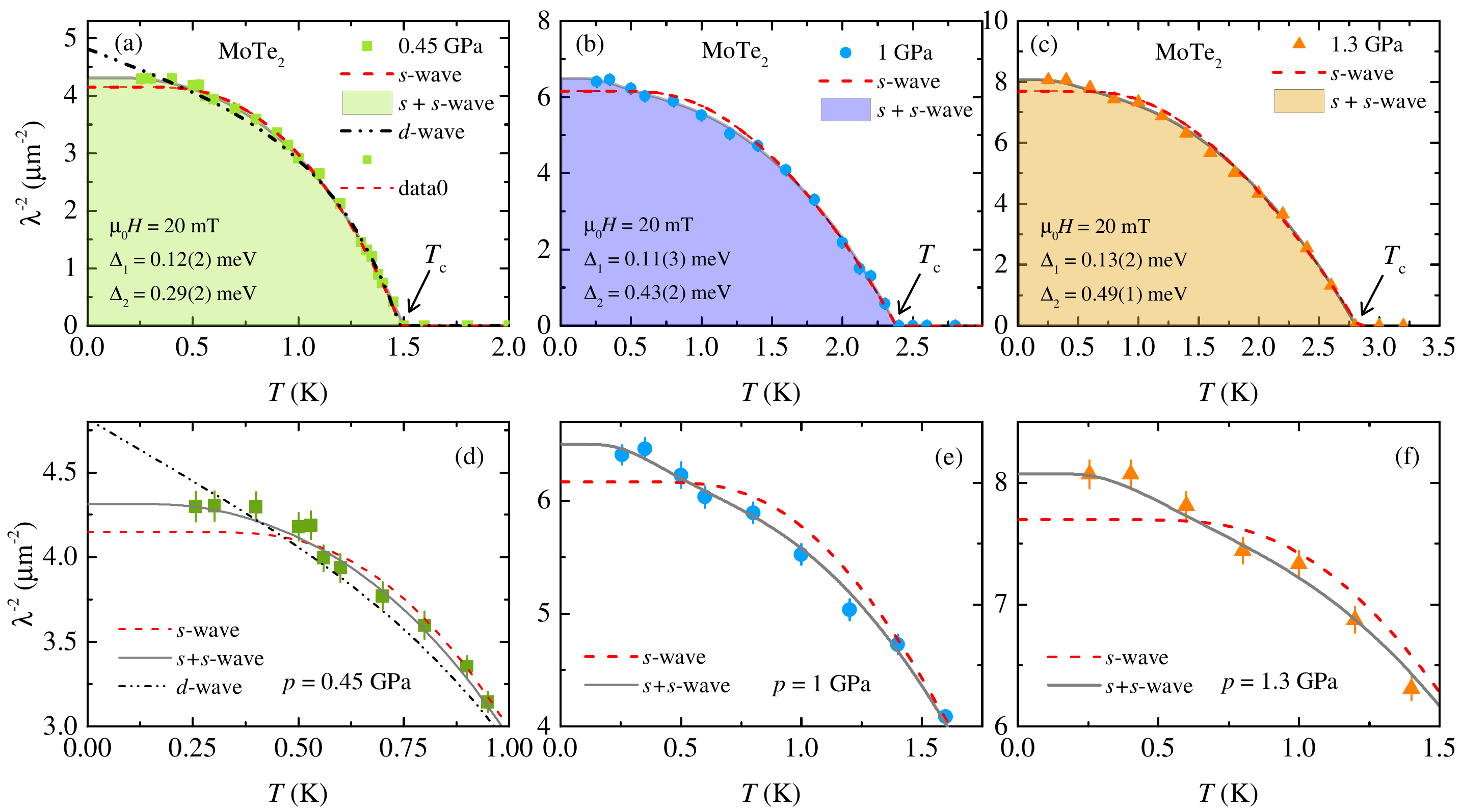}
\vspace{-0.8cm}
\caption{ (Color online)  \textbf{Pressure evolution of the ${\lambda}^{-2}(T)$.}
The temperature dependence of ${\lambda}^{-2}$ measured at various applied hydrostatic pressures for MoTe$_{2}$. The solid lines correspond to a two-gap ($s+s$)-wave model, the dashed and the dotted lines represent a fit using a single gap $s$- and $d$-wave models, respectively. }
\label{fig4}
\end{figure*}
 
  The temperature dependence of the muon spin depolarization rate ${\sigma}_{{\rm sc}}$,
which is proportional to the second moment of the field distribution (see Method section), of MoTe$_{2}$ in the SC state at selected pressures is shown in Fig.~ \ref{fig3}. Below $T_{{\rm c}}$ the relaxation rate ${\sigma}_{{\rm sc}}$ starts to increase from zero with decreasing temperature due to the formation
of the FLL. There is a significant increase of the low-temperature value ${\sigma}_{{\rm sc}}$(0.25
K) as well as $T_{{\rm c}}$ under pressure (see Fig.~\ref{fig3}): ${\sigma}_{{\rm sc}}$(5
K) increases by a factor of ${\sim}$ 2 from $p$ = 0 GPa to p = 1.3 GPa.
Interestingly, at all pressures the form of the temperature dependence of ${\sigma}_{{\rm sc}}$,
which reflects the topology of the SC gap, shows slight upturn at ${\sim}$ 700 mK with indication of 
the saturation upon further lowering the temperature below ${\sim}$ 700 mK. We show in the following how these behaviours indicate the presence of the
two isotropic $s$-wave gaps on the Fermi surface of MoTe$_{2}$.

 In order to investigate the symmetry of the SC gap, we note that ${\lambda}(T)$ is related to the relaxation rate
${\sigma}_{{\rm sc}}(T)$ by the equation \cite{Brandt}: 
\begin{equation}
\frac{\sigma_{sc}(T)}{\gamma_{\mu}}=0.06091\frac{\Phi_{0}}{\lambda^{2}(T)},
\end{equation}
where ${\gamma_{\mu}}$ is the gyromagnetic ratio of the muon, and
${\Phi}_{{\rm 0}}$ is the magnetic-flux quantum. Thus, the flat $T$-dependence
of ${\sigma}_{{\rm sc}}$ observed at various pressures for low temperatures
(see Fig.~\ref{fig3}) is consistent with a nodeless superconductor, in which
$\lambda^{-2}\left(T\right)$ reaches its zero-temperature value exponentially.

To proceed with a quantitative analysis, we consider the local (London)
approximation (${\lambda}$ ${\gg}$ ${\xi}$, where ${\xi}$ is the
coherence length) and employ the empirical ${\alpha}$-model.
The model, widely used in previous investigations of the penetration
depth of multi-band superconductors \cite{Bastian,Tinkham,carrington,padamsee,GuguchiaPRB},
assumes that the gaps occuring in different bands, besides a common
$T_{{\rm c}}$, are independent of each other. The superfluid
density is calculated for each component separately \cite{GuguchiaPRB}
and added together with a weighting factor. For our purposes, a two-band
model suffices, yielding: 
\begin{equation}
\frac{\lambda^{-2}(T)}{\lambda^{-2}(0)}=\omega_{1}\frac{\lambda^{-2}(T,\Delta_{0,1})}{\lambda^{-2}(0,\Delta_{0,1})}+\omega_{2}\frac{\lambda^{-2}(T,\Delta_{0,2})}{\lambda^{-2}(0,\Delta_{0,2})},
\end{equation}
where ${\lambda}(0)$ is the penetration depth at zero temperature,
${\Delta_{0,i}}$ is the value of the $i$-th SC gap ($i=1$, 2) at
$T=0$~K, and ${\omega}_{i}$ is the weighting factor which measures
their relative contributions to ${\lambda^{-2}}$ (i.e. ${\omega}_{1}+{\omega}_{2}=1$).

\begin{figure*}[t!]
\centering
\includegraphics[width=1.0\linewidth]{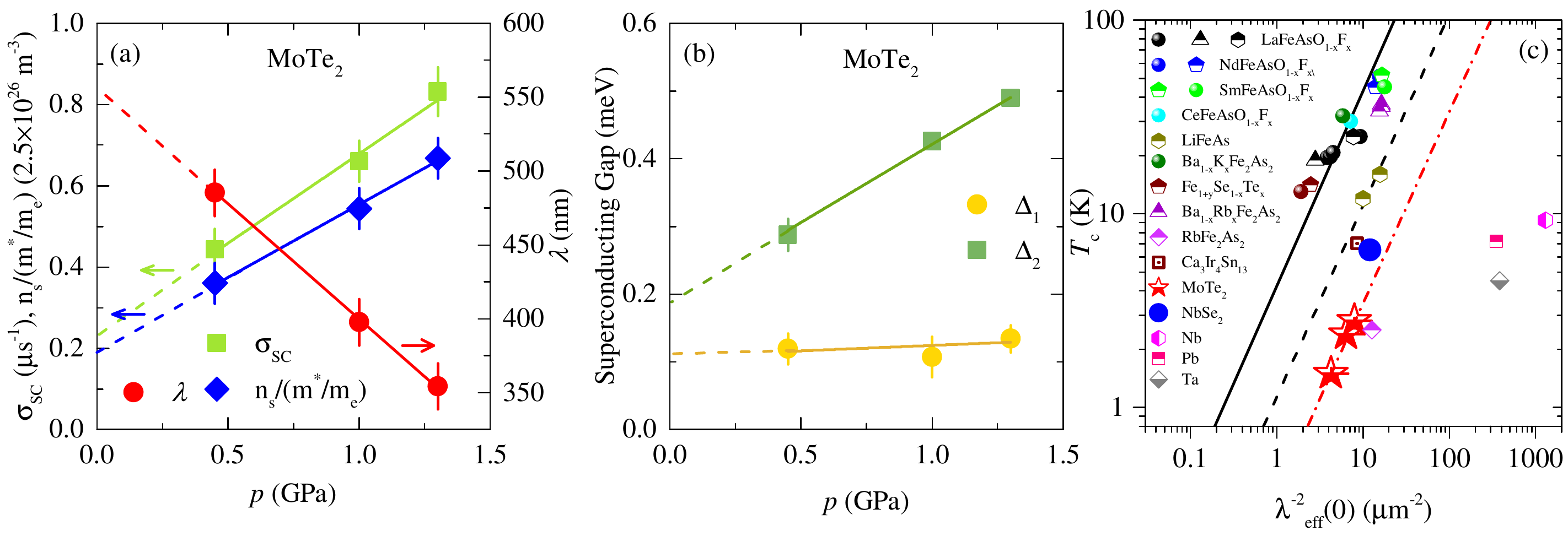}
\vspace{-0.7cm}
\caption{ (Color online) \textbf{Pressure evolution of various quantities.} The SC muon depolarisation rate ${\sigma}_{SC}$, magnetic penetration depth ${\lambda}$ and the superfluid density $n_{s}/(m^{*}m_{e}$ (a) as well as the zero-tempearture gap values ${\Delta}_{1,2}$(0) (b) as a function of hydrostatic pressure. The dashed lines are the guides to the eyes and the solid lines represent the linear fits to the data. (c) A plot of $T_{\rm c}$ against the ${\lambda}^{-2}(0)$ obtained from our ${\mu}$SR experiments in 
MoTe$_{2}$. The dashed red line represents the linear fit to the MoTe$_{2}$ data. Uemura plot for various cuprate and Fe-based HTSs are shown 
(Refs. \cite{GuguchiaPRB,Shermadini,Luetkens,Kim,Takeshita,Carlo,Khasanov2008,Khasanov2010,Pratt}).  The relation observed for underdoped cuprates is also shown (solid line for hole doping \cite{Uemura1,Uemura3,Uemura4,Uemura5,Uemura6} and dashed black line for electron doping \cite{Shengelaya}). The points for various conventional BCS superconductors and for NbSe$_{2}$ are also shown.}
\label{fig5}
\end{figure*}

 The results of this analysis are presented in Fig. \ref{fig4}a-c, where the
temperature dependence of ${\lambda^{-2}}$ for MoTe$_{2}$
is plotted at various pressures. We consider two different possibilities
for the gap functions: either a constant gap, $\Delta_{0,i}=\Delta_{i}$,
or an angle-dependent gap of the form $\Delta_{0,i}=\Delta_{i}\cos2\varphi$,
where $\varphi$ is the polar angle around the Fermi surface. 
The dashed and the solid lines represent a fit to the data using a $s$-wave
and a $s$ + $s$-wave models, respectively. The analysis appears
to rule out the simple $s$-wave model as an adequate
description of ${\lambda^{-2}}$($T$) for MoTe$_{2}$. 
The two gap $s$ + $s$-wave scenario with a small gap ${\Delta}_{1}$ ${\simeq}$ 0.12(3) meV  and a large gap ${\Delta}_{2}$ (with the pressure independent weighting factor of ${\omega}_{2}$ = 0.87), describes the experimental data remarkably well. A $d$-wave gap symmetry was also tested, shown with black dotted line in Fig. \ref{fig4}a, but was found to be inconsistent with the data. This conclusion is supported by a ${\chi}^{2}$ test, revealing a 
${\chi}^{2}$ for the $d$-wave model by ${\sim}$ 30 ${\%}$ higher than the one for ($s+s$)-wave model for $p$ = 0.45 GPa. The pressure dependence of all the parameters extracted from the data
analysis within the ${\alpha}$ model are plotted in Figs.~\ref{fig5}(a-b).
From Fig.~\ref{fig5}a a substantial decrease of ${\lambda}\left(0\right)$ (increase of ${\sigma}_{{\rm sc}}$)
with pressure is evident. At the maximum applied pressure of $p$
= 1.3 GPa the reduction of ${\lambda}\left(0\right)$ is approximately
25 ${\%}$ compared to the value at $p$ = 0.45 GPa. The small gap ${\Delta}_{1}$ ${\simeq}$ 0.12(3) meV
stays nearly unchanged by pressure, while the large gap ${\Delta}_{2}$ increases from ${\Delta}_{2}$ ${\simeq}$ 0.29(1) meV at $p$ = 0.45 GPa to ${\Delta}_{2}$ ${\simeq}$ 0.49(1) meV at $p$ = 1.3 GPa, i.e., by approximately 70 ${\%}$.

 In general, the penetration depth ${\lambda}$ is given as a function of $n_{\rm s}$, $m^{*}$, ${\xi}$ and the mean free path $l$ as
 
\begin{equation}
\begin{aligned}
\frac{1}{\lambda^2} = \frac{4\pi n_se^2}{m^*c^2} \times  \frac{1}{1 + \xi/l}, 
\end{aligned}
\end{equation}  
 
For systems close to the clean limit, ${\xi}$/$l$ ${\rightarrow}$ 0, the second term essentially becomes unity, and the simple relation 1/${\lambda}^{2}$ $\propto$ $n_{s}/m^{*}$ holds. Considering the $H_{c2}$ values of MoTe$_{2}$, reported in Ref.\cite{QiCava} we estimated ${\xi}$ ${\simeq}$ 26 nm and 14 nm for $p$ = 0.45 GPa and 1 GPa, respectively.  At ambient pressure, the in-plane mean free path $l$ was estimated to be $l$ ${\simeq}$ 100 - 200 nm \cite{Luo}. No estimates are currently available for $l$ under pressure. However, in-plane $l$ is most probably independent on pressure, considering the fact that the effect of compression is mostly interlayered (i.e., intralayer Mo-Te bond length is almost not changed, especially in our investigated pressure region), thanks to the unique anisotropy of the van der Waals structure. Thus, in view of the short coherence length and relatively large $l$, we can assume that MoTe$_{2}$ lies close to the clean limit \cite{Frandsen}. With this assumption, we obtain the ground-state value  $n_{s}/(m^{*}/m_{e}$) ${\simeq}$ 0.9 ${\times}$ 10$^{26}$ m$^{-3}$, 1.36 ${\times}$ 10$^{26}$ m$^{-3}$, and 1.67 ${\times}$ 10$^{26}$ m$^{-3}$ for $p$ = 0.45 GPa, 1 GPa, and 1.3 GPa respectively. Interestingly, $n_{s}/(m^{*}/m_{e}$) increases substantially under pressure and this will be discussed below.

  One of the essential findings of this paper is the observation of two-gap superconductivity in $T_{d}$-MoTe$_{2}$. Recent ARPES \cite{Liang} experiments on MoTe$_{2}$ revealed the presence of three bulk hole pockets (a circular hole pocket around the Brillouin zone center and the two butterfly-like hole pockets) and two bulk electron pockets, that are symmetrically distributed along the ${\Gamma}$-X direction with respect to the Brillouin zone center ${\Gamma}$. Since several bands cross the Fermi surface in MoTe$_{2}$, two-gap superconductivity can be understood by assuming that the SC gaps open at two distinct types of bands. Now the interesting question arises: How consistent is the observed two-gap superconductivity with the possible topological nature of superconductivity in $T_{d}$-MoTe$_{2}$? Note that the superconductor $T_{d}$-MoTe$_{2}$ represents a case of  time-reversal-invariant Weyl semimetals, which has broken inversion symmetry. Recently, the microscopic interactions and the SC gap symmetry in a time-reversal-invariant topological superconductivity in Weyl semimetals was studied based on the fluctuation-exchange approach \cite{Hosur}. In a Weyl semimetal, the FSs carry nonzero Chern numbers of the Berry`s phase gauge field. A simple formula \cite{Qi}, relates the FS Chern number to the topological invariant ${\nu}$ for a time-invariant TSC, considering a set of FSs with Chern numbers {$C_{\rm j}$}:

\begin{equation}
\begin{aligned}
{\nu} = \frac{1}{2} {\sum_{j \in FS}}C_{\rm j}{sgn}({\Delta}_{j}), 
\label{eq3}
\end{aligned}
\end{equation}  

where $\Delta_{\rm j}$ is the pairing gap function on the $j$-th FS. 
A TSC is implied by ${\nu} {\neq}$  0. It was shown that for isotropic Weyl nodes (\cite{Hosur}), the superconducting gap function projected onto the FSs in any time-symmetric gapped phase must have the form $\Delta_{\rm j}$(k) = $\Delta_{\rm j}$, i.e., it must be independent of \textbf{k}. Using further theoretical considerations \cite{Hosur}, the topological invariant defined in Eq. \ref{eq3} is expressed as

\begin{equation}
\begin{aligned}
{\nu} = sgn({\Delta}_{1}) - sgn ({\Delta}_{2}). 
\end{aligned}
\end{equation}  
 
 Therefore the necessary condition for the TSC is that ${\Delta}_{\rm 1}$ and ${\Delta}_{\rm 2}$ have opposite signs. 
So, for TSC the gaps can be momentum independent on each FS but must alternate
in sign between different FSs. As suggested in Ref.  \cite{Hosur}, for the TSC to be favoured over the trivial state, in which ${\Delta}_{\rm 1}$ and ${\Delta}_{\rm 2}$  have the same sign, the useful conditions are Coulomb repulsion as well as attractive interactions, such as those mediated by phonons, and a fine-tuning of their relative strengths.  In this way, the net interaction can be made to change sign over large momenta, comparable to the separation of the Weyl nodes. It was also suggested \cite{Hosur} that strong disorder may be more detrimental to topological superconductivity than to ordinary $s$-wave superconductivity due to inter-FS scattering processes.  
${\mu}$SR experiments alone can not distinguish between $s^{+-}$ (topological) and $s^{++}$ (trivial) pairing states. However, considering the strong  suppression of $T_{\rm c}$  in MoTe$_{2}$ by disorder \cite{Balicas1,Daniel}, we suggest that $s^{+-}$ state is more likely to be realized than the $s^{++}$ state. Further phase sensitive experiments are desirable  to distinguish between $s^{+-}$ and $s^{++}$ states in MoTe$_{2}$.

 Besides the two-gap superconductivity, another interesting observation is the strong enhancement of the superfluid density ${\lambda}^{-2}(0)$ ${\propto}$ $n_{s}/(m^{*}/m_{e}$) and its linear scaling with $T_{c}$ (see Fig. 5c). 
$n_{s}/(m^{*}/m_{e}$) increases by factor of ${\sim}$ 1.8 between $p$ = 0.45 GPa and 1.3 GPa. 
Note that our DFT calculations also show the strong changes in the electronic band structure  (see Supplementary Note I and Supplementary Figure 6). The nearly linear relationship between $T_{\rm c}$ and the superfluid density was first noticed in hole-doped cuprates in 1988-89 \cite{Uemura1,Uemura3}, and possible relevance to the crossover from Bose Einstein Condensation to BCS condensation have been discussed in several subsequent papers \cite{Uemura4,Uemura5,Uemura6}. The linear relationship was noticed mainly with systems following
the line of the ratio between  $T_{\rm c}$ and their effective Fermi temperature
$T_{\rm F}$ being about  $T_{\rm c}$/ $T_{\rm F}$ ${\sim}$ 0.05, which means about 4-5 times
reduction of  $T_{\rm c}$ from the ideal Bose Condensation temperature for
a non-interacting Bose gas composed of the same number of Fermions pairing
without changing their effective masses.  The present results on MoTe$_{2}$ and
NbSe$_{2}$ \cite{Le} in Figure 5c demonstrate that a linear relation holds for these
systems but with the ratio  $T_{\rm c}$/$T_{\rm F}$ being reduced further by a factor
of 16-20.  It was also noticed \cite{Shengelaya} that electron-doped cuprates
follow another line with their $T_{\rm c}$/$T_{\rm F}$ reduced by a factor of ${\sim}$ 4
from the line of hole doped cuprates. Since the present system MoTe$_{2}$
and NbSe$_{2}$ falls into the clean limit, the linear relation is unrelated to
pair breaking, and can be regarded to hold between $T_{\rm c}$ and $n_{s}/m^{*}$.

 In a naive picture of BEC to BCS crossover, systems with small
$T_{\rm c}$/$T_{\rm F}$ (large $T_{\rm F}$) are considered to be in the "BCS" side, while the
linear relationship between $T_{\rm c}$ and $T_{\rm F}$ is expected only in the BEC side.
Figure 5c indicates that the BEC-like linear relationship
may exist in systems with $T_{\rm c}$/$T_{\rm F}$ reduced by a factor 4 to 20
from the ratio in hole doped cuprates, presenting a new challenge for
theoretical explanations.

 In conclusion, we provide the first microscopic investigation of the superconductivity in $T_{d}$-MoTe$_{2}$. Specifically, the zero-temperature magnetic penetration depth ${\lambda}\left(0\right)$
and the temperature dependence of ${\lambda^{-2}}$ were studied in the Type-II Weyl-semimetal $T_{d}$-MoTe$_{2}$ by means of ${\mu}$SR experiments as a function of pressure up to p ${\simeq}$ 1.3 GPa. Remarkably, the temperature dependence of $1/\lambda^{2}\left(T\right)$  is inconsistent with the presence of nodes in the gap as well as with a simple isotropic $s$-wave type of the order  parameter. However, it is well described by a two-gap ($s+s$) - wave scenario, indicating multigap superconductivity in MoTe$_{2}$.  
We also excluded time reversal symmetry breaking in the high-pressure SC state with sensitive zero-field ${\mu}$SR experiments, clasifying MoTe$_{2}$ as time-reversal-invariant superconductor with broken inversion symmetry. 
In this type of superconductor, a two-gap ($s+s$)-wave model is consistent with a topologically non-trivial superconducting state if the gaps ${\Delta}_{\rm 1}$ and ${\Delta}_{\rm 2}$, at different Fermi surfaces, have opposite signs. ${\mu}$SR experiments alone can not distinguish between sign changing $s^{+-}$ (topological) and $s^{++}$ (trivial) pairing states. However, considering the previous report on the strong  suppression of $T_{\rm c}$  in MoTe$_{2}$ by disorder, we suggest that $s^{+-}$ state is more likely to be realized in MoTe$_{2}$ than the $s^{++}$ state. Should $s^{+-}$ be the SC gap symmetry, the high pressure state of MoTe$_{2}$ is, to our knowledge, the first known example of a Weyl superconductor, as well as the first example of a time reversal invariant topological superconductor. 
Finally, we observed a linear correlation between $T_{c}$ and the zero-temperature superfluid density ${\lambda^{-2}}(0)$ in MoTe$_{2}$, which together with the observed two-gap behaviour points to the unconventional nature of superconductivity in $T_{d}$-MoTe$_{2}$. We hope, the present results will stimulate theoretical investigations to obtain a microscopic understanding of the relation between superconductivity and the topologically non-trivial electronic structure of $T_{d}$-MoTe$_{2}$.

\section{METHODS}

\textbf{Sample preparation}: High quality single crystals and polycrystalline samples were obtained by mixing of molybdenum foil (99.95 ${\%}$) and tellurium lumps (99.999+${\%}$) in a ratio of 1:20 in a quartz tube and sealed under vacuum. The reagents were heated to 1000$^{\rm o}$C within 10 h. They dwelled at this temperature for 24 h, before they were cooled to 900$^{\rm o}$C within 30 h (polycrystalline sample) or 100 h (single crystals). At 900$^{\rm o}$C the tellurium flux was spined-off and the samples were quenched in air. The obtained MoTe$_{2}$ samples were annealed at 400$^{\rm o}$C for 12 h to remove any residual tellurium.\\ 

\textbf{Pressure cell}:  Pressures up to 1.3 GPa were generated in a single wall piston-cylinder
type of cell made of CuBe material, especially designed to perform ${\mu}$SR experiments under
pressure \cite{GuguchiaPressure,Andreica}. As a pressure transmitting medium Daphne oil was used. The pressure was measured by tracking the SC transition of a very small indium plate by AC susceptibility. The filling factor of the pressure cell was maximized. The fraction of the muons stopping in the sample was approximately 40 ${\%}$.\\

\textbf{${\mu}$SR experiment}: 

 In a ${\mu}$SR experiment nearly 100 ${\%}$ spin-polarized muons ${\mu}$$^{+}$
are implanted into the sample one at a time. The positively
charged ${\mu}$$^{+}$ thermalize at interstitial lattice sites, where they
act as magnetic microprobes. In a magnetic material the 
muon spin precesses in the local field $B_{\rm \mu}$ at the
muon site with the Larmor frequency ${\nu}_{\rm \mu}$ = $\gamma_{\rm \mu}$/(2${\pi})$$B_{\rm \mu}$ (muon
gyromagnetic ratio $\gamma_{\rm \mu}$/(2${\pi}$) = 135.5 MHz T$^{-1}$). 
Using the $\mu$SR technique important length scales of superconductors can be measured, namely the magnetic penetration depth $\lambda$ and the coherence length $\xi$. If a type II superconductor is cooled below $T_{\rm c}$ in an applied magnetic field ranged between the lower ($H_{c1}$) and the upper ($H_{c2}$) critical fields, a vortex lattice is formed which in general is incommensurate with the crystal lattice with vortex cores separated by much larger distances than those of the unit cell. Because the implanted muons stop at given crystallographic sites, they will randomly probe the field distribution of the vortex lattice. Such measurements need to be performed in a field applied perpendicular to the initial muon spin polarization (so called TF configuration). 

 ${\mu}$SR experiments under pressure were performed at the ${\mu}$E1 beamline of the Paul Scherrer Institute (Villigen, Switzerland, where an intense high-energy ($p_{\mu}$ = 100 MeV/c) beam of muons is implanted in the sample through the pressure cell. The low background GPS (${\pi}$M3 beamline) and low-temperature LTF instruments were used to study the single crystalline as well as the polycrystalline samples of MoTe$_{2}$ at ambient pressure.\\
  
\textbf{Analysis of TF-${\mu}$SR data}: 

 The TF ${\mu}$SR data were analyzed by using the following functional form:\cite{Bastian}
\begin{equation}
\begin{aligned}
P(t)=A_s\exp\Big[-\frac{(\sigma_{sc}^2+\sigma_{nm}^2)t^2}{2}\Big]\cos(\gamma_{\mu}B_{int,s}t+\varphi) \\
 + A_{pc}\exp\Big[-\frac{\sigma_{pc}^2t^2}{2}\Big]\cos(\gamma_{\mu}B_{int,pc}t+\varphi), 
\end{aligned}
\end{equation}
 Here $A_{\rm s}$ and $A_{\rm pc}$  denote the initial assymmetries of the sample and the pressure cell, respectively. $\gamma/(2{\pi})\simeq 135.5$~MHz/T 
is the muon gyromagnetic ratio, ${\varphi}$ is the initial phase of the muon-spin ensemble and $B_{\rm int}$ represents the
internal magnetic field at the muon site. The relaxation rates ${\sigma}_{\rm sc}$ 
and ${\sigma}_{\rm nm}$ characterize the damping due to the formation of the FLL in the SC state and of the nuclear 
magnetic dipolar contribution, respectively. In the analysis ${\sigma}_{\rm nm}$ was assumed 
to be constant over the entire temperature range and was fixed to the value obtained above 
$T_{\rm c}$ where only nuclear magnetic moments contribute to the muon depolarization rate ${\sigma}$.
The Gaussian relaxation rate, ${\sigma}_{\rm pc}$, reflects the depolarization due
to the nuclear moments of the pressure cell. 
The width of the pressure cell signal increases below $T_{c}$. As shown previously \cite{MaisuradzePC}, this is due to 
the influence of the diamagnetic moment of the SC sample on the pressure cell, leading to the temperature dependent
${\sigma}_{\rm pc}$ below $T_{c}$. In order to consider this influence we assume the linear coupling between ${\sigma}_{\rm pc}$ and the field shift of the internal magnetic field in the SC state: 
${\sigma}_{\rm pc}$($T$) = ${\sigma}_{\rm pc}$($T$ ${\textgreater}$ $T_{\rm c}$) + $C(T)$(${\mu}_{\rm 0}$$H_{\rm int,NS}$ - ${\mu}_{\rm 0}$$H_{\rm int,SC}$), where  ${\sigma}_{\rm pc}$($T$ ${\textgreater}$ $T_{\rm c}$) = 0.25 ${\mu}$$s^{-1}$ is the temperature independent Gaussian relaxation rate. ${\mu}_{\rm 0}$$H_{\rm int,NS}$ and ${\mu}_{\rm 0}$$H_{\rm int,SC}$ are the internal magnetic fields measured in the normal and in the SC state, respectively. 
As indicated by the solid lines in Figs.~2b,c the ${\mu}$SR data are well described by Eq.~(1).
The good agreement between the fits and the data demonstrates that the model used 
describes the data rather well.\\

\textbf{Analysis of ${\lambda}(T)$}:



 ${\lambda}$($T$) was calculated within the local (London) approximation (${\lambda}$ ${\gg}$ ${\xi}$) 
by the following expression \cite{Bastian,Tinkham}:
\begin{equation}
\frac{\lambda^{-2}(T,\Delta_{0,i})}{\lambda^{-2}(0,\Delta_{0,i})}=
1+\frac{1}{\pi}\int_{0}^{2\pi}\int_{\Delta(_{T,\varphi})}^{\infty}(\frac{\partial f}{\partial E})\frac{EdEd\varphi}{\sqrt{E^2-\Delta_i(T,\varphi)^2}},
\end{equation}
where $f=[1+\exp(E/k_{\rm B}T)]^{-1}$ is the Fermi function, ${\varphi}$ is the angle along the Fermi surface, and ${\Delta}_{i}(T,{\varphi})={\Delta}_{0,i}{\Gamma}(T/T_{\rm c})g({\varphi}$)
(${\Delta}_{0,i}$ is the maximum gap value at $T=0$). 
The temperature dependence of the gap is approximated by the expression 
${\Gamma}(T/T_{\rm c})=\tanh{\{}1.82[1.018(T_{\rm c}/T-1)]^{0.51}{\}}$,\cite{carrington} 
while $g({\varphi}$) describes 
the angular dependence of the gap and it is replaced by 1 for both an $s$-wave and an $s$+$s$-wave gap,
and ${\mid}\cos(2{\varphi}){\mid}$ for a $d$-wave gap.

\section{Acknowledgments}~
The ${\mu}$SR experiments were carried out at the Swiss Muon Source (S${\mu}$S) Paul Scherrer Insitute, Villigen, Switzerland. X-ray PDF measurements were conducted on beamline 28-ID-2 of the National Synchrotron Light Source II, a U.S. Department of Energy (DOE) Office of Science User Facility operated for the DOE Office of Science by Brookhaven National Laboratory under Contract No. DE-SC0012704. Z. Guguchia gratefully acknowledges the financial support by the Swiss National Science Foundation (SNFfellowship P2ZHP2-161980). The material preparation at Princeton was supported by the Gordon and Betty Moore Foundation EPiQS initiative, grant GBMF-4412. Z.G. and Y.J.U. thank Prof. Andrew Millis for useful discussions. Work at Department of Physics of Columbia University is supported by US NSF DMR-1436095 (DMREF) and NSF DMR-1610633. Work in the Billinge group was supported by U.S. Department of Energy, Office of Science, Office of Basic Energy Sciences (DOE-BES) under contract No. DE-SC00112704. S. Banerjee acknowledges support from the National Defense Science and Engineering Graduate Fellowship program. A.S. acknowledges support from the SCOPES grant No. SCOPES IZ74Z0-160484. 






\newpage

\section{SUPPLEMENTAL MATERIAL \\ }


\subsection{I. DFT calculations of the electronic band structure}

  We used van der Waals density (vdW) functional and the projector augmented wave (PAW) method \cite{PAW},
as implemented in the VASP code \cite{VASP}. We adopted the generalized gradient approximation (GGA) proposed by Perdew \emph{et al} (PBE) \cite{PBE} and DFT-D2 vdW functional proposed by Grimme \emph{et al} \cite{Grimme,Harl} as a nonlocal correlation. Spin-orbit coupling (SOC) is included in all cases.
A plane wave basis with a kinetic energy cutoff of 500 eV was employed. We used a $\Gamma$-centered \textbf{k}-point mesh of 15$\times$9$\times$5. Optimized lattice parameters of $T_d$ phase are $a$=3.507, $b$=6.371, and $c$=13.743\AA, close to the previous experimental values; $(a,b,c)$=(3.468, 6.310, 13.861) \cite{Tamai} and (3.458, 6.304,13.859) \cite{Wang}.  

  Figure 6 shows the calculated band structure of $T_{d}$-MoTe$_{2}$ at ambient $p$ and for $p$ = 1.3 GPa, showing the strong changes in 
the electronic band structure, induced by pressure. This is consistent with the strong pressure dependence of $n_{s}/m^{*}$, observed experimentally.

\subsection{II. Confirmation of the orthorhombic ($T_{d}$) to monoclinic ($T^{'}$) phase transition in MoTe$_{2}$}

Total scattering measurements were carried out on the XPD (28-ID-2) beamline at the National Synchrotron Light Source II (NSLS-II), Brookhaven National Laboratory. A finely ground powder of MoTe$_2$ was prepared in an inert Argon chamber, and sealed in 1.02mm (OD) polyimide capillaries. Diffraction patterns were collected in a Debye-Scherrer geometry with an X-ray energy of 67.127 keV ($\lambda=0.18470$~\AA) using a large-area 2D PerkinElmer detector (2048$^2$ pixels with 200 $\mu m^{2}$ pixel size). The detector was mounted with a sample-to-detector distance of 344.79 mm, to achieve a balance between $q$-resolution and $q$-range. The sample was measured at 100K and 300K using an Oxford CS-700 cryostream, allowing ample time for the material to thermalize. The experimental geometry, $2\theta$ range, and detector misorientations were calibrated by measuring a crystalline nickel powder directly prior to data collection at each temperature point, with the experimental geometry parameters refined using the PyFAI program~\cite{kieffer_pyfai_2013}.

Raw 2D diffraction patterns are azimuthally integrated to obtain the 1D scattering intensity $I(Q)$ where $Q=4\pi\sin\theta/\lambda$. Standardized corrections are then made to the data and $I(Q)$ is normalized by dividing by the total scattering cross-section of the sample, resulting in the structure function $S(Q)$
\begin{equation}
\label{eq:SQ}
  S(Q) - 1 = \frac{I_{c}(Q) - \sum{c_i\abs{f_{i}(Q)^2}}}{\sum{c_i\abs{f_{i}(Q)^2}}}
\end{equation}
where $I_{c}$ is the coherent scattering intensity as a function of the momentum transfer $Q$ and $c_i$ and $f_i$ are the concentration and X-ray scattering factor, per atom type $i$. Because the scattering cross section becomes very small at high-$Q$, an important result of this normalization is that the high angle data are significantly amplified \cite{egami_underneath_2003}. This is illustrated in Figure 1(a,b) where we show the standard integrated XRD patterns $I(Q)$ versus the reduced total scattering structure function $F(Q)$ (insets), where $F(Q) = Q[S(Q) - 1]$. While the raw diffraction pattern appears to have almost no distinguishable peaks near the upper limit of the shortened $Q$-range, the form factor normalized $F(Q)$ contains well resolved, high amplitude peaks extending to a $Q$-range twice the maximum value shown for $I(Q)$. The high degree of structural coherence observed for MoTe$_2$ is in contrast to diffraction studies of similar layered TMDs, where diffuse scattering from turbostratic disorder and other defects often dominates the high angle signal\cite{petkov_application_2000}. 

The full $Q$-range where structure can be resolved, including both Bragg and diffuse components, is shown in the insets of Figure 7. The structure function is Fourier transformed to obtain the pair distribution function (PDF), using PDFgetX3~\cite{juhas_pdfgetx3:_2013} within xPDFsuite~\cite{yang_xpdfsuite:_2014}. The experimental PDF, $G(r)$, is the truncated Fourier transform of $F(Q)$
\begin{equation}
\label{eq:FTofSQtoGr}
  G(r) = \frac{2}{\pi}
          \int_{\qmin}^{\qmax} F(Q)\sin(Qr) \: \dd Q,
\end{equation}
For both MoTe$_2$ measurements (100K and 300K), the $Q_{max}$ was chosen to be 22.5~\AA$^{-1}$ to give the best tradeoff between statistical noise and real-space resolution.   

The PDF is a histogram of interatomic distances that gives the probability of finding pairs of atoms in a material, separated by a distance $r$. $G(r)$ can be calculated from a structure model according to
\begin{align}
\label{eq:Grfromrhor}
  G(r) &= 4 \pi r \left[ \rho(r) - \rho_{0} \right], \\
  \rho(r) &= \frac{1}{4 \pi r^{2} N}
            \sum_{i}\sum_{j \neq i}
                \frac{f_{i}f_{j}}{\langle f \rangle ^{2}}
                \delta (r - r_{ij}). \nonumber
\end{align}
Here, $\rho_{0}$ is the atomic number density of the material and $\rho(r)$ is
the atomic pair density, which is the mean weighted density of neighbor atoms at
distance $r$ from an atom at the origin. The sums in $\rho(r)$ run over all
atoms ($N$) in the model, with periodic boundary conditions. $f_{i}$ is the scattering factor of atom $i$, $\langle f\rangle$ is the average scattering factor and $r_{ij}$ is the distance between atoms $i$ and $j$. \eqs{Grfromrhor} are used to fit the PDF calculated from a model to the experimentally measured PDFs. The \pdfgui program was used to construct unit cells from reference structures, carry out the refinements, and determine the agreement between simulated PDFs and data, quantified by the residual function $R_w$ \cite{farrow_pdffit2_2007}.

We included three candidate structure models for local structure refinements of MoTe$_2$ PDFs measured at 100K and 300K. The monoclinic 1T$'$-MoTe$_2$ (SG: $P2_{1}/m$) structure reported by B.E. Brown \cite{brown_crystal_1966}, the orthorhombic 1T$_d$-MoTe$_2$ (SG: $Pmn2_1$) structure, also referred to as the $\gamma$ form, given in Wang \etal \cite{Wang}, and a hexagonal 2H polytype (SG:$P6_{3}/mmc$) reported by D. Puotinen and R.E. Newnham \cite{puotinen_crystal_1961}. The 2H structure was ruled out as both 1T structures gave significantly better fits for both low-temperature and high-temperature datasets. The T$'$ and T$_d$ PDF refinements were performed conservatively by adjusting just a few parameters in the structure models: lattice ($a$,$b$,$c$), scale, delta-2 (a parameter for correlated motion effects), one isotropic atomic displacement parameter (ADP) for the 4 Mo atoms, and one ADP for the 8 Te atoms, with both structures containing 12 atoms in the unit cell. For the monoclinic phase, an additional symmetry allowed $\beta$ angle was refined. Atomic positions were not refined. The instrumental resolution parameters $Q_{damp}=0.0217$ and $Q_{broad}=0.009$ were determined through refinements of the nickel calibrant, and kept fixed for all other refinements.

 PDFs for MoTe$_2$ are refined over an  $r$-range from $1.5<r<15$~\AA, slightly larger than the $c$-axis length for both T$'$ and T$_d$ models. The PDF at low-$r$ contains structural information from the full $F(Q)$ range and as a result, refinements are very sensitive to deviations from the average structure. This is advantageous when characterizing the local atomic environment for MoTe$_2$, with both candidate models containing layers of edge-sharing, distorted MoTe$_6$ octahedra (Figure 7) within the refinement range selected. The approach is different from traditional powder XRD Rietveld refinements of similar materials, where a substantial amount of diffuse scattering is neglected, and phase identification is restricted to a small number of low angle Bragg reflections, thus increasing the number of possible structures that may fit equivalently to the same data. 

The results of the PDF analysis are summarized in Figure 8. For MoTe$_2$ measured at 100K there is a clear improvement in the agreement factor ($R_w$) when the data is refined to the orthorhombic T$_d$ model, with smaller residuals indicating better fits. The difference in refinement quality is more significant above the orthorhombic to monoclinic phase transition ($T_s~\sim~250K$) and the 300K PDF is in good agreement with the T$'$ model. By varying a small number of structural parameters we ensure that these differences in $R_w$ accurately represent a structural transformation. Additionally, the refined lattice constants differ minimally from the starting values provided in the reference T$'$ and T$_d$ models. The best-fit PDF at 100K using the T$_d$ structure, yields refined lattice parameters of $a=3.474$\AA, $b=6.328$\AA, $c=13.884$\AA, and small ADPs, $U_{iso}$(Mo)$=0.0043$; $U_{iso}$(Te)$=0.0056$. For MoTe$_2$ at 300K, the monoclinic structure refines $a=6.335$\AA, $b=3.471$\AA, $c=13.853$\AA; $\beta=93.886 \si{\degree}$ and  $U_{iso}$(Mo)$=0.0058$; $U_{iso}$(Te)$=0.0064$. 

In Figure 9, we index the Bragg profile to support the results from the more quantitative PDF refinements. Here we simply simply compare the raw diffraction patterns measured at 100K and 300K to powder diffraction patterns calculated from the candidate models using the program VESTA \cite{momma_vesta:_2008}.  We plot the data over a $2\theta$ range where the T$'$ and T$_d$ models have Bragg peaks that are well resolved from one another to highlight distinguishable features that differ between the phases in the measured XRD pattern.

\clearpage

\begin{onecolumngrid}

\begin{figure}[t!]
\includegraphics[width=0.8\linewidth]{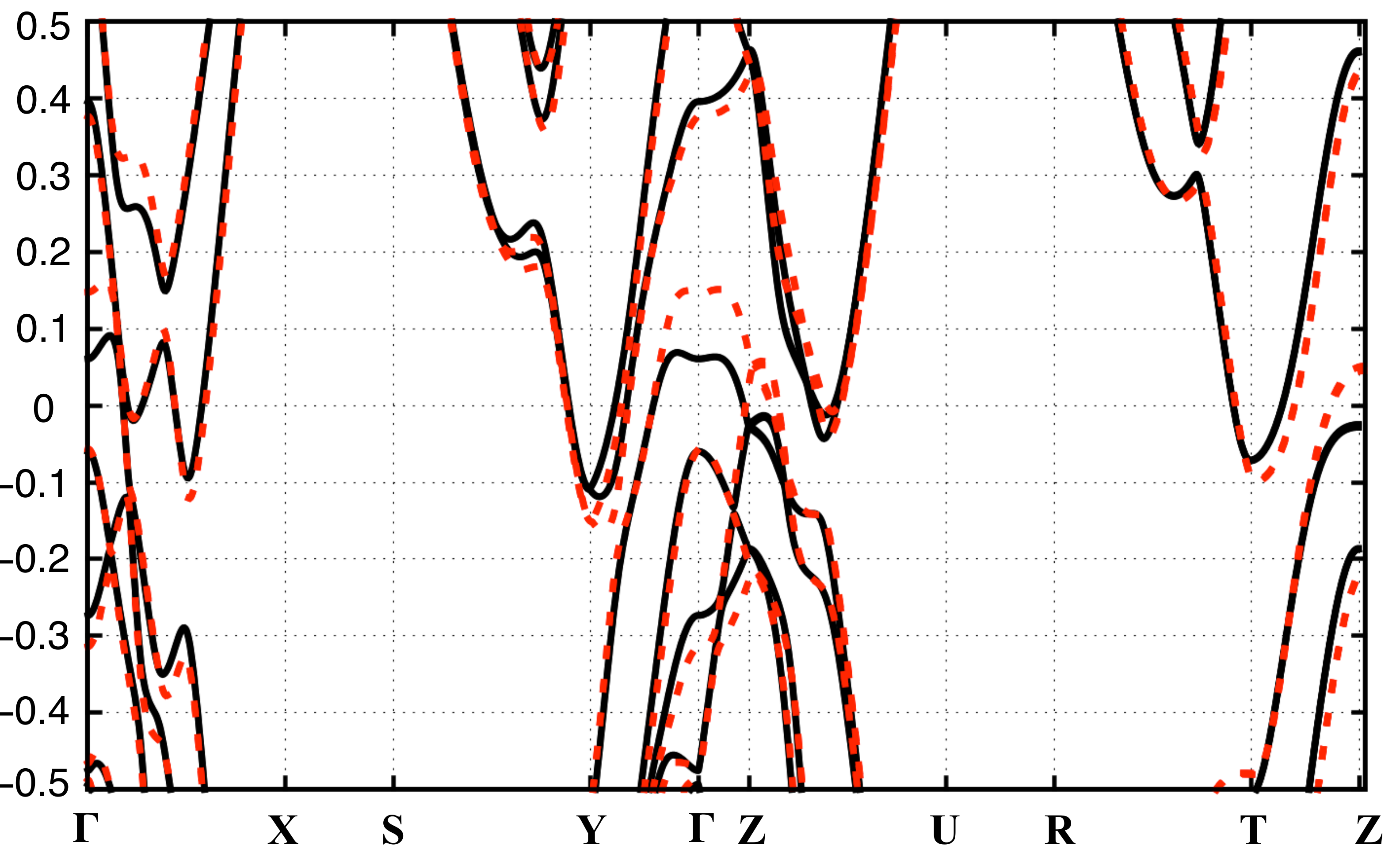}
\caption{ (Color online) Band structure of $T_{d}$-MoTe$_{2}$ at ambient $p$ (solid black curves) and for $p$ = 1.3 GPa (dashed red curves).}
\label{fig1}
\end{figure}

\begin{figure}[tbh]
\includegraphics[width=1\columnwidth]{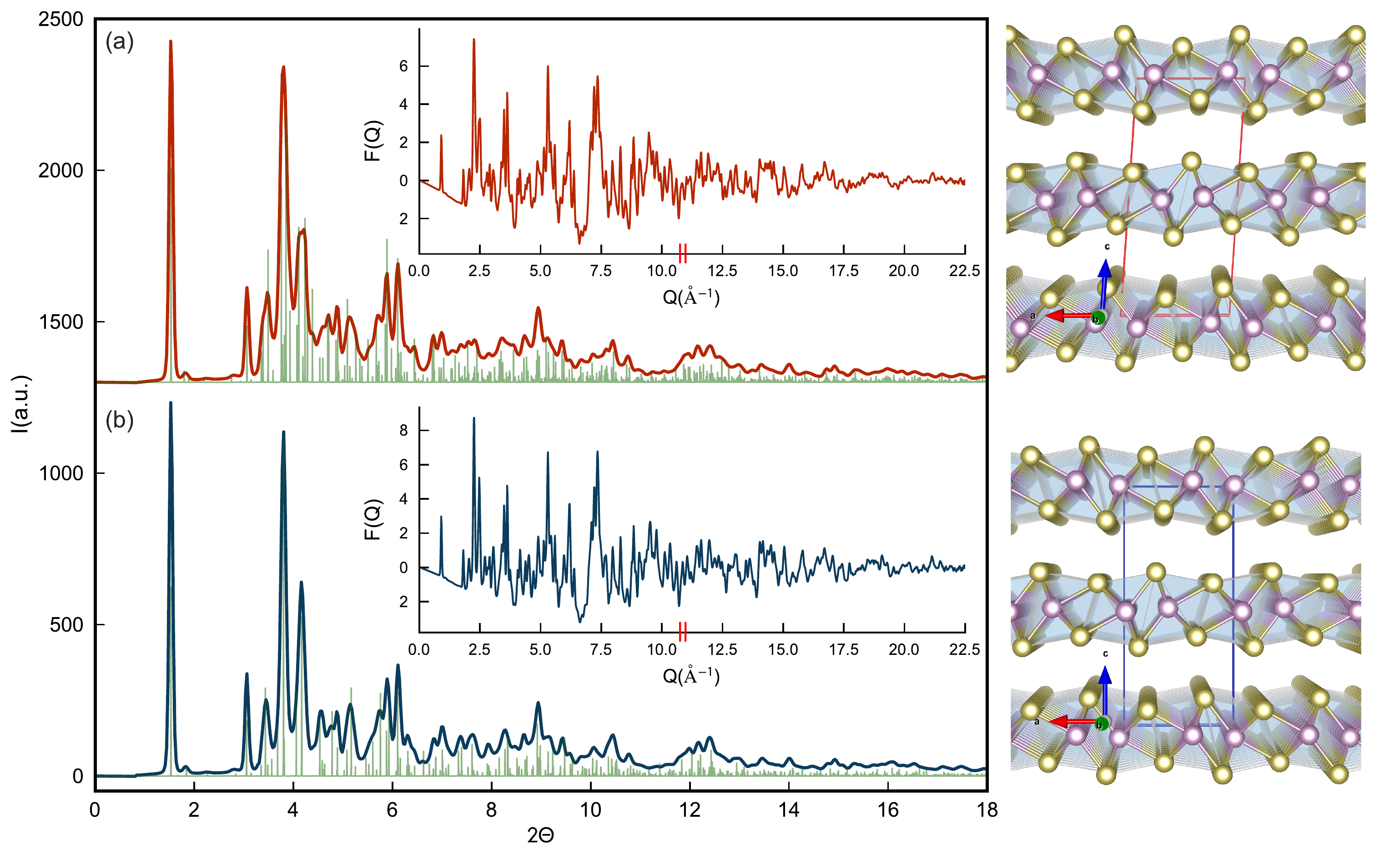}
\caption{\label{xrd-1}Powder diffraction patterns of MoTe$_2$ measured at 300K (a) and 100K (b) plotted as a function of $Q$ and truncated where Bragg peaks are no longer distinguishable. The insets show the total scattering structure function $F(Q)$ where the full range used for the PDF transformation is provided. Small vertical lines in red mark the upper limit of $I(Q)$ given in (a,b), past which a significant amount of scattered intensity is seen in the structure function. Fine green lines correspond to calculated Bragg peaks from candidate models T$'$ (a) and T$_d$ (b). The differences between the Bragg reflections are inspected over a smaller angular range in Figure 9. To the right we show the centrosymmetric monoclinic structure plotted above the non-centrosymmetric orthorhombic structure, from references discussed in the text. The MoTe$_6$ octahedral units are shaded in blue. Note that in order to provide a useful projection to compare T$'$ and T$_d$, the unit cell orientation for the orthorhombic structure is changed from the standard (abc) setting for $Pmn2_1$ to (ba-c) for the symmetry equivalent $Pnm2_1$ space group.}
\end{figure}

\begin{figure}[tbh]
\includegraphics[width=1\columnwidth]{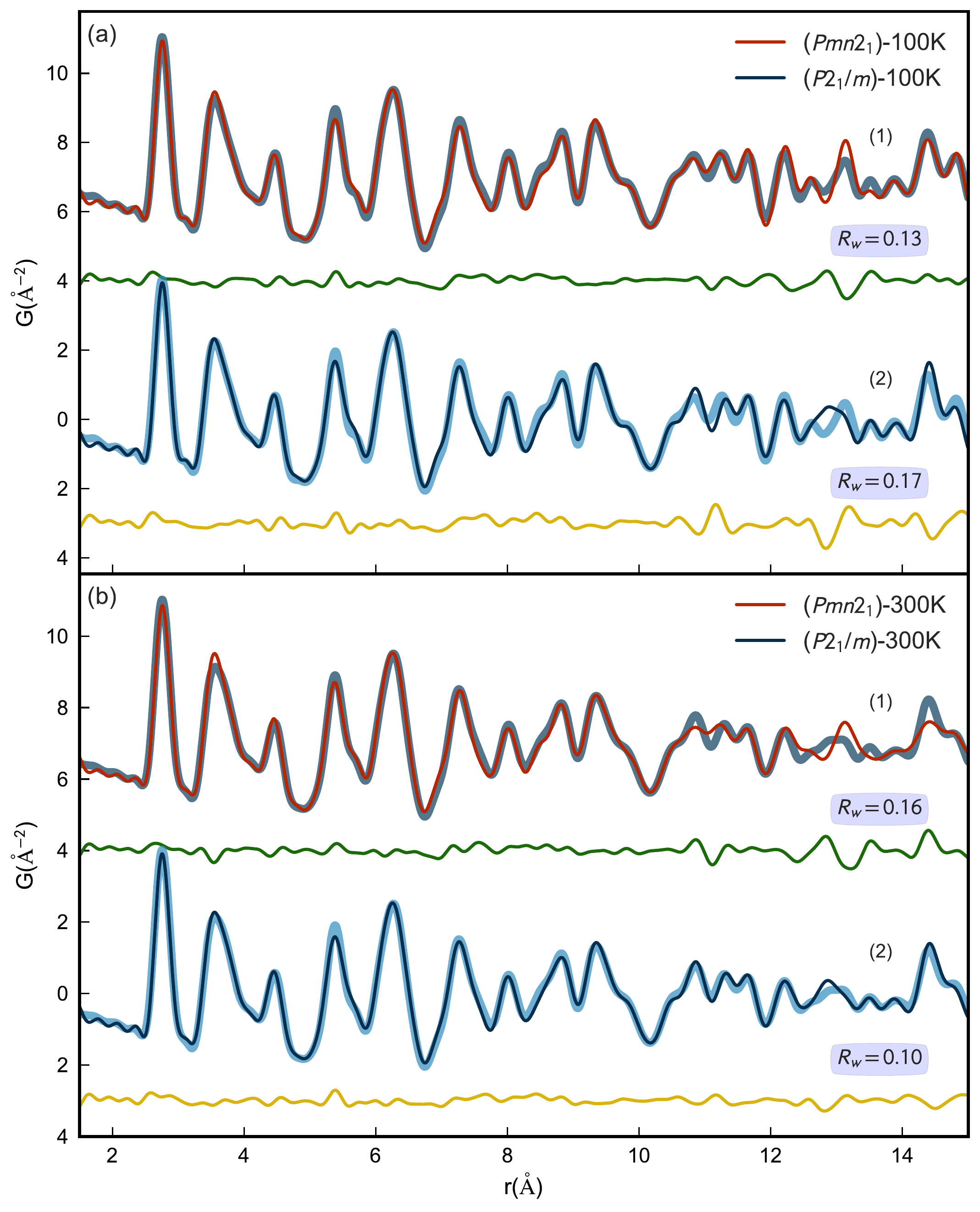}
\caption{\label{pdf-1} (a) PDF local structure refinements of MoTe$_2$ measured at 100K fit to the T$_d$ structure (1) and the T$'$ structure (2). (b) an analogous comparison for the 300K measurement. Thick curves are the raw PDF data, and the calculated PDF from model is overlayed as thin solid lines, and labeled in the legend. Offset difference curves in green and yellow are simply the calculated PDF subtracted from the measured PDF. The goodness-of-fit parameter ($R_w$) is provided for each of the four refinements, which confirm that the 100K data is better described as T$_d$ and the 300K data as T$'$.}
\end{figure}

\begin{figure}[tbh]
\includegraphics[width=1\columnwidth]{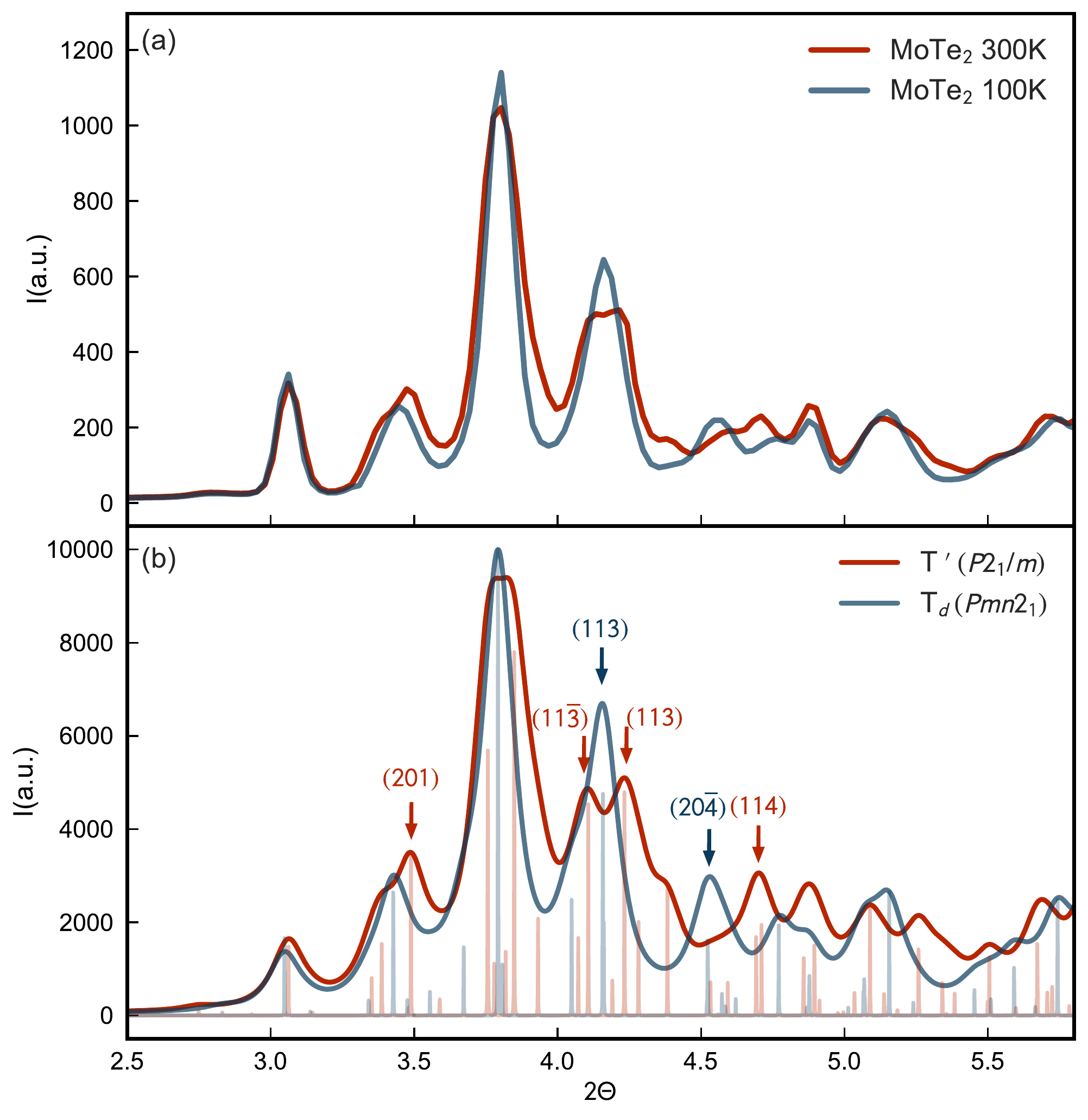}
\caption{\label{xrd-2} (a) Raw diffraction patterns for 300K and 100K measurements using an incident wavelength of $\lambda=0.18470$~\AA. The $2\theta$ range shown here corresponds to the $Q$-range between 1.48 and 3.44 \AA$^{-1}$ in Figure 7. (b) Bragg profile calculated from unmodified candidate structures. Solid curves in red and blue are the Bragg peaks calculated from the reflections (sharp lines), broadened uniformly with an FWHM=0.1 in $2\theta$\;(=0.059\AA$^{-1}$ in Q). The orthorhombic $(113)$ reflection measured at 100K, splits into the $(11\overline{3})$ and $(113)$ at 300K, which was also reported for WTe$_2$ \cite{lu_origin_2016}}
\end{figure}

\end{onecolumngrid}

\clearpage

\end{document}